# Cross-Issue Solidarity and Truth Convergence in Opinion Dynamics

Zong Xuan Tan[1] and Kang Hao Cheong[2]

[1]Yale University, New Haven, CT 06520, United States
[2]Engineering Cluster, Singapore Institute of Technology, 10 Dover Drive, S138683, Singapore

**Abstract**

How do movements and coalitions which engage with multiple social issues succeed in cross-issue solidarity, and when do they instead become fragmented? To address this, the mechanisms of cross-issue interaction have to be understood. Prior work on opinion dynamics and political disagreement has focused on single-issue consensus and polarization. Inspired by practices of cross-issue movement building, we have developed a general model of multi-issue opinion dynamics where agreement on one issue can promote greater inclusivity in discussing other issues, thereby avoiding the pitfalls of exclusivist interaction, where individuals engage only if they agree sufficiently on every issue considered. Our model shows that as more issues come into play, consensus and solidarity can only be maintained if inclusivity towards differing positions is increased. We further investigate whether greater inclusivity and compromise across issues lead people towards or away from normative truth, thereby addressing concerns about the non-ideal nature of political consensus.

# 1 Introduction

In the wake of rising populism in liberal democracies, and a spate of unexpected electoral outcomes in the United States and the United Kingdom, a flurry of commentary has emerged, linking the cause of these political upheavals to increasing polarization and fragmentation.[1–3] While concerns about fragmentation are not new, and have been diagnosed as a problem for both left-wing and right-wing coalitions,[4–6] recent criticism in the US has been directed at the purported divisiveness of leftist identity politics, as well as the related practice of cross-issue intersectional activism.[7] Critics of these practices charge that they cause fragmentation by simultaneously tackling multiple social issues, which they believe creates more room for disagreement.[8–10] Defenders argue instead that they are necessary to engender change that is genuinely inclusive.[11–13] Underlying these arguments are different premises about the dynamics of political disagreement when multiple issues are involved — does fragmentation become more likely as more issues are involved, or can cross-issue interaction instead give rise to solidarity and consensus?

One approach to addressing this question is through opinion dynamics, an inter-disciplinary field that has attracted the likes of physicists, mathematicians, and engineers. Numerous models and applications have been developed in connection with opinion dynamics,[14–18] and some of its predictions have been validated by theory and experimentation in social psychology.[19–22] While the mathematical approach of opinion dynamics necessarily reduces the complex semantics of opinions to either numbers[23–25] or yes/no variables,[26] the phenomena captured by these models are still of great richness and diversity, ranging from consensus formation[27] to rumour spreading[28–30] and innovation diffusion.[31] Of prominence are the bounded confidence models of opinion dynamics, which stipulate that agents only consider the opinions of others if those opinions are close enough to their own — i.e., their confidence is bounded. Such models were first developed by Hegselmann and Krause,[23] as well as Deffuant and Weisbuch,[24,25] and they are notable because they readily demonstrate the emergence of political polarization and fragmentation under the assumption of bounded confidence. Their psychological basis can be found in social judgement theory (SJT), which predicts that people's opinions move closer together only if they fall within each others' 'latitude of acceptance'.[19,32]

Although many variations upon bounded confidence dynamics have since been developed, they have primarily been limited to how opinions on a single issue evolve.[33] Crucial to our present investigation, however, is the role of cross-issue interaction. Such interactions can be highly significant and warrant detailed investigation — single-issue models predict that sharp disagreement on one issue will lead to disengagement between individuals, but this neglects the possibility that they might still hear each other out because they share some other belief. Conversely, it may be that two individuals are unable to come to terms on an issue that they mostly agree about, simply because they sharply disagree on some other issue that drives a wedge between them. Of the limited work that has been done on multi-issue opinion dynamics, some studies have extended the concept of one-dimensional latitudes of acceptance into $n$-dimensional "spheres",[33] addressed cases where opinions are about appropriate budget allocations,[34,35] or focused on how agreement on one issue might induce tolerance (but not agreement) on others.[36] Other work, while not concerned with cross-issue interaction directly, has taken such interaction into account while explaining phenomena like the difference between perceived and actual levels of political polarization.[37] The rich potential of cross-issue interactions, where similarity along one issue might promote agreement along another, remains under-explored.

In contrast to previous approaches, the present study seeks to develop models which help elucidate the conditions under which cross-issue mobilization leads to either consensus or fragmentation, an investigation that bears directly upon the debates discussed earlier. In doing so, we address the following two questions of relevance: Firstly, how does increasing the number of issues affect the likelihood of cross-issue consensus, and how might cohesion be maintained regardless? Secondly, given that a common worry about consensus across political divides is that it may simply lead to false compromises, and that the space of 'correct opinions' may grow increasingly small as more issues come into play, what is the relationship between pursuing cross-issue consensus and ensuring convergence to normative truth?

To answer these questions, we developed different possible models of cross-issue interaction. Some of these were less inclusive, in that individuals would only engage if they agreed sufficiently on all considered issues, while others were more inclusive, in that agreement on a single issue would increase tolerance for disagreement on others. The effects of such inclusivity upon consensus were explored in tandem with the effects of the number of issues. To investigate the relationship between truth convergence and consensus, we modified the forgoing models to incorporate interactions between individuals and the truth, and also considered the role of the truth's extremity. In doing so, we have produced comprehensive models which capture the collective effects of these factors upon cross-issue solidarity and truth convergence.

## 2 Models of opinion dynamics

We first introduce some notation before showing the development of our models from previous work. Consider a group of $n$ individuals, where each individual $i$ holds opinions on $m$ different issues. These opinions are represented as a vector $x^i$ of length $m$, where the $k$th-dimensional component of the vector (i.e. $i$'s opinion on the $k$th issue) is given by a real number $x_k^i$.

Each component $x_i^k$ is assumed to lie within a finite range, say $[0, 1]$. That is, the most extreme position in support of a particular issue is assigned a value of $1$, and the most extreme position against it is assigned a value of $0$. For each dimension $k$, we also define the population opinion vector, $x_k := (x_k^1, x_k^2, ..., x_k^n)$, which contains all the group's opinions on the $k$th issue. Opinions evolve with time, so we denote the opinion vector at time $t$ as $x_k(t)$. When $m = 1$, we omit the subscript and write $x(t) \equiv x_1(t)$.

## 2.1 Bounded confidence in a single dimension

To account for the limited acceptance of other opinions and the possibility of polarization predicted by social judgement theory, we build on the bounded confidence model first proposed by Hegselmann & Krause (2002).[23] In bounded confidence models, individuals only give consideration to others who share opinions sufficiently similar to their own. After each round of discussion, some individuals who used to interact with each other may find that their opinions are now too far apart, and begin to ignore each other. Predictably, these dynamics result in polarization if individuals only take into consideration opinions that are very similar to their own.

Formally, the model is defined by assigning each individual $i$ a neighborhood of acceptance $I(i, x)$. This neighborhood contains all individuals with sufficiently similar opinions

$$I(i, x) := \{j : |x^j - x^i| \leq \epsilon\} \tag{1}$$

where $\epsilon$ is called the latitude of acceptance. Each individual updates their opinion simply by averaging over all the opinions which fall within the neighborhood of acceptance, giving us

$$x^i(t+1) = |I(i,x)|^{-1} \sum_{j \in I(i,x)} x^j(t) \tag{2}$$

Here, $|I(i, x)|$ denotes the number of individuals inside $I(i, x)$.

## 2.2 Bounded confidence in multiple dimensions

We proceed to extend the above model into multiple dimensions, thereby accounting for the possibility of interaction between opinions on several issues. We start with the case where issues are completely independent and do not influence each other. Following that, we draw upon differing practices of multi-issue advocacy and movement building to present some possible modes of cross-issue interaction.

The first of these is *exclusivist interaction*, where individuals only engage with others if they agree sufficiently on *all* issues under consideration. The second is *inclusivist interaction*, where individuals engage with others as long as they agree sufficiently on *at least one* issue under consideration. We then present a general model of cross-issue interaction which interpolates between these two extremes using a parameter we term the *degree of inclusivity*.

**Independent issues**

The most straightforward extension to multiple dimensions is to treat all issues as *independent*, i.e., opinions on different issues do not influence each other at all. For each dimension $k$, we thus have the same dynamics as before:

$$I_k(i, x) := \{j : |x_k^j - x_k^i| \leq \epsilon_k\} \tag{3}$$

$$x_k^i(t+1) = |I_k(i,x)|^{-1} \sum_{j \in I_k(i,x)} x_k^j(t) \tag{4}$$

$I_k(i, x)$ and $\epsilon_k$ are respectively the neighborhood and latitude of acceptance in dimension $k$, $1 \leq k \leq m$. Since the dynamics in each dimension can be analyzed separately, results from the one-dimensional case carry over.

**Exclusivist interaction**

Opinions and beliefs are rarely completely independent. Instead, they often influence one another, and our perceptions of others' beliefs. What happens if two individuals largely agree on one issue, but disagree sharply on another? One possibility is that they refuse to engage with each other. Though seemingly close-minded, there are circumstances under which most would find this reasonable. For example, if a vegetarian meets an animal lover who also happens to endorse slavery, most of us would think the vegetarian justified in refusing engagement, despite whatever gains they might make for the cause of animal rights by befriending the pro-slavery animal lover.

This mode of cross-issue interaction can be described as *exclusivist* in character, because each individual only gives weight to others if their opinions are sufficiently close in *all* dimensions, and excludes everyone else. We can thus define the neighborhood of acceptance $I_\cap(i, x)$, and the corresponding update rule:

$$I_\cap(i, x) := \{j : |x_k^j - x_k^i| \leq \epsilon_k \text{ for all } k\} \tag{5}$$

$$x_k^i(t+1) = |I_\cap(i, x)|^{-1} \sum_{j \in I_\cap(i,x)} x_k^j(t) \tag{6}$$

Note that $I_\cap(i, x) = \cap_{k=1}^m I_k(i, x)$. That is, $I_\cap(i, x)$ is the intersection of all the neighborhoods $I_k(i, x)$ defined in the model with independent issues. Given the exclusivity of these dynamics, it is expected the chances for multi-issue consensus will decrease as more issues come under consideration, with opinions fragmenting into large numbers of non-interacting clusters instead.

**Inclusivist interaction**

Returning to the scenario where two individuals are in agreement on some issues but not others, the obvious alternative is for them to continue engaging despite their disagreements. In the most extreme case, they would engage so long there is *at least one* issue upon which they agree. This might occur, for example, if the belief agreed upon is something fundamental to both of their social identities, such as religion or a commitment to one's nation, leading to willingness to compromise on other issues. Alternatively, such compromise could just be strategic – both individuals might decide they stand more to gain from engaging on the issue where they share similar opinions, even though this might mean letting go of their disagreements on other issues.

Since this mode of cross-issue interaction emphasizes inclusion in dialogue despite disagreement, it can be described as *inclusivist* in character. The neighborhood of acceptance $I_\cup(i, x)$ and the corresponding update rule are defined as

$$I_\cup(i, x) := \{j : \text{there exists } k \text{ where } |x_k^j - x_k^i| \leq \epsilon_k\} \tag{7}$$

$$x_k^i(t+1) = |I_\cup(i,x)|^{-1} \sum_{j \in I_\cup(i,x)} x_k^j(t) \tag{8}$$

Note that $I_\cup(i,x) = \cup_{k=1}^m I_k(i,x)$. That is, $I_\cup(i,x)$ is the union of all the neighborhoods $I_k(i,x)$. Naturally, we might expect these dynamics to have a greater chance of multi-issue consensus than the exclusivist case, and to maintain a high chance of consensus even when the number of issues increases.

**Generalized inclusivity**

Neither the exclusivist nor inclusivist mode of interaction completely captures the complexity of cross-issue interactions. In general, it is reasonable to suppose that agreement on one issue will only foster agreement on another issue to a limited extent. To that end, we can define a model that interpolates between the two extremes. Intuitively speaking, the general model says that if individuals agree sufficiently on one issue, they become more tolerant of disagreements on other issues. For example, if someone loves classical music but hates pop, they might give consideration to the opinions of a fellow classical music lover who is slightly more fond of pop music, as opposed to the opinions of someone who is similarly fond of pop, but does not like classical music at all.

A more algorithmic explanation is as follows. During each round of discussion, an individual first checks to see if an interlocutor's opinion is within their latitude of acceptance $\epsilon_k$ for at least one issue $k$. If this is true, they then have an *expanded* latitude of acceptance $\alpha_k \epsilon_l$ for every other issue $l \neq k$, where $\alpha_k \geq 1$ is a factor called the *degree of inclusivity*. If, for each of these issues $l \neq k$, the opinions held by the interlocutor are within the expanded latitude of acceptance (i.e. $|x_l^j - x_l^i| \leq \alpha_k \epsilon_l$ for all $l \neq k$), the individual will give their interlocutor some weight of consideration. Following this mechanism, we can define the neighborhood of acceptance and the corresponding update rule:

$$I_k^*(i,x) := \{j : |x_k^j - x_k^i| \leq \epsilon_k \text{ and } |x_l^j - x_l^i| \leq \alpha_k \epsilon_l \text{ for all } l \neq k\} \tag{9}$$

$$I^*(i,x) := \bigcup_{k=1}^m I_k^*(i,x) \tag{10}$$

$$x_k^i(t+1) = |I^*(i,x)|^{-1} \sum_{j \in I^*(i,x)} x_k^j(t) \tag{11}$$

Here, $I_k^*(i,x)$ corresponds to the expanded neighborhood of acceptance of an individual $i$ for all opinions that fall within the latitude of acceptance for the $k$th issue (i.e., all $x^j$ where $|x_k^j - x_k^i| \leq \epsilon_k$). Since we have one expanded neighborhood for each issue $k$, we take the union of all of them to obtain the overall neighborhood of acceptance, $I^*(i,x)$.

The role played by $\alpha_k$ is crucial for the generality of this model. It can be understood as the degree by which agreement on the $k$th issue makes one more inclusive of differences on all other issues. If we set $\alpha_k = 1$ for all $k$, then all issues are minimally inclusive with respect to each other, and we get the exclusivist model described earlier. But if we set $\alpha_k = \infty$ for all $k$, this is maximally inclusive, and we get the inclusivist model from before. A prediction that follows is that the degree of consensus should increase with the values of $\alpha_k$, while opinion fragmentation decreases. Depending on how large $\alpha_k$ is, the degree of consensus may decrease with the number of issues (as predicted for the exclusivist case), or it may be maintained (as predicted for the inclusivist case).

Figure 1 provides a visual representation of the models described thus far. In each sub-figure, the plus sign marks the position of an individual in two-dimensional opinion space, with $x_1$ and $x_2$ denoting opinions on the first and second issues respectively. Dashed lines surround the neighborhoods of acceptance for the marked individual. Figure 1a shows the neighborhood in the one-dimensional case — i.e. when individuals only consider the first issue during interactions. Any opinion within the vertical strip is acceptable, because it falls within $\epsilon_1 = 0.1$ of the individual's opinion on issue 1. Figure 1b depicts the exclusivist case, which has a rectangular neighborhood because opinions have to fall within $\epsilon_k$ of all issues $k = 1, 2$ to be acceptable. Figure 1c depicts the inclusivist case. The neighborhood is a cross which extends to the borders of the opinion space, because falling within $\epsilon_k$ for either issue $k$ is acceptable. Figure 1d shows the general case when $\alpha_1 = \alpha_2 = 2.5$. It should be noted that this neighborhood interpolates between the extremes of inclusivity and exclusivity, appearing as cross with limited bar-span along each dimension.

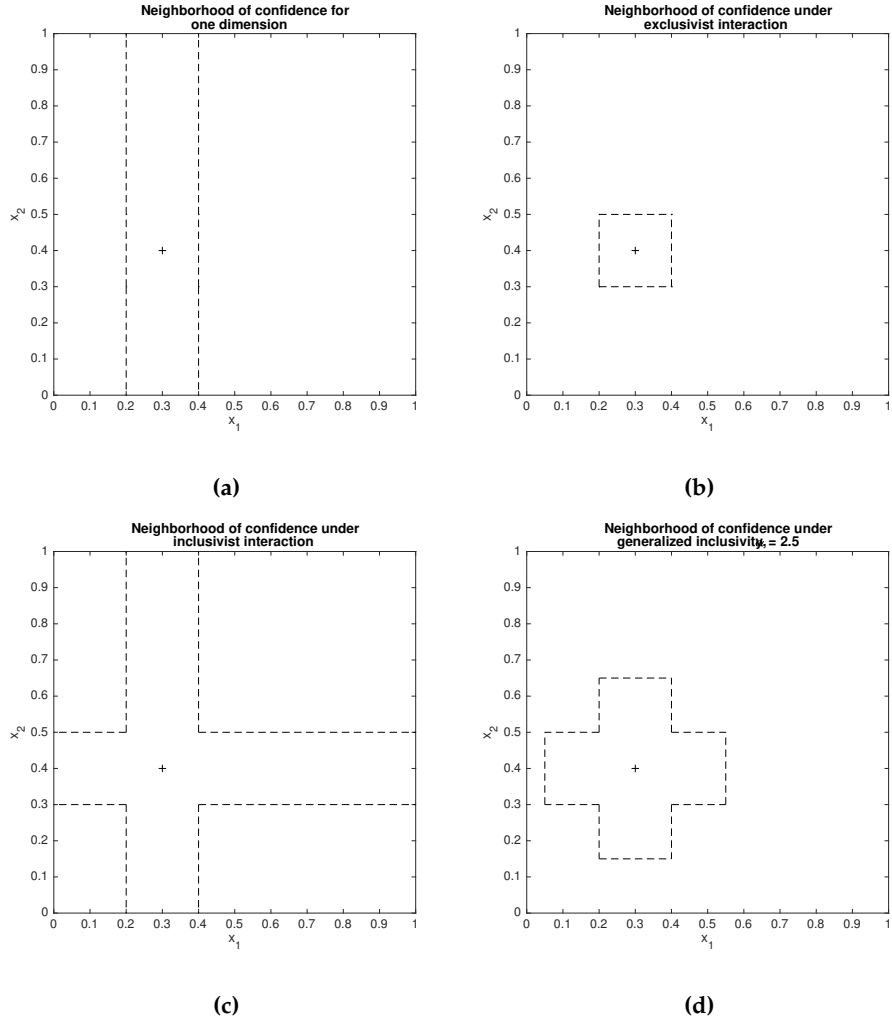

**Figure 1:** Neighborhoods of acceptance (dashed boundaries) about the point $(0.3, 0.4)$, with latitudes of acceptance $\epsilon_k = 0.1$ for all $k$, and $\alpha_k = 2.5$ for all $k$ in (d). The number of dimensions $m = 1$ in (a), but the second dimension is plotted for ease of comparison.

## 2.3 Dynamics in the presence of truth

A common worry about political compromise is that the consensus reached may not be ideal or true, i.e. the result of a fallacious "authority of the many" or "argument to moderation" rather than careful reasoning.[38,39] This can be used as justification against inclusive dialogue or consensus, because if a party strongly believes that they are in the right, they might also believe that it is better to stand one's ground than let go of what they view as the truth. Such behavior is espe-cially pertinent in the context of multi-issue deliberation, because with more issues to "stand one's ground" upon, fragmentation is even more likely to occur.

However, is it necessarily the case that consensus will lead people away from the truth? Here we propose an extension to the above models that allows this question to be investigated. Suppose that for each issue $k$, there does in fact exist a "correct opinion" or "normative truth" $T$, with the coordinates $x_k = T_k$. Suppose also that individuals close in opinion to the truth will discover or interact with it in some way, perhaps by investigating the world or thinking more deeply about the issue. An elegant way to include this in a bounded confidence model of opinion dynamics is to model the truth as as a virtual 'individual' $x^0$ with a fixed location in opinion space, $x^0 \equiv T = (T_1, ..., T_m)$. This 'individual' $x^0 \equiv T$ never changes its position, but all individuals nearby to $T$ in opinion space will include it in their neighborhoods of acceptance. They will therefore update their opinions in response to the truth, and if there are no other individuals pulling them in the opposite direction, their opinions will eventually converge towards $T$, resulting in consensus that is also true.

Undoubtedly, such a desired result will not always occur, depending on how inclusive or exclusive people are, and how many people's opinions are initially situated close to the truth. We investigated the conditions under which convergence towards the truth occurs by varying the parameters of the models described above, the results of which are presented below.

## 3 Methods

The models described above were implemented in MATLAB R2017 (MathWorks). At every time step, the transition matrix for each opinion dimension was calculated as a function of the population's opinions (represented as a matrix), and then used to compute the population's opinions in the next time step. The number of final opinion clusters was computed using the *unique* function in MATLAB, thereby extracting the distinct points in the opinion space to which opinions eventually converged. Convergence to the truth was determined by checking whether an individual's opinion was within $\delta = 0.025$ of the truth's location by the final time step of each simulation (exact convergence to the truth is not possible because the truth's location is fixed). In determining aggregate trends, all simulations were repeated at least 100 times, with the results averaged. The generality of these trends were corroborated by analytical derivations, which are included in Section 1.1 of the Supplementary Information.

## 4 Results

### 4.1 Effects of inclusivity and dimensionality on consensus

Each of the multi-dimensional models described above was simulated with the same latitudes of acceptance $\epsilon_k$, in order to investigate their respective propensities towards consensus or fragmentation. Figures 2 to 5 below show sample runs for each model when the number of issues was $m = 2$, with the initial opinions were distributed uniformly at random over the space of possible opinions. $m = 2$ was chosen to allow visualization of the evolution of opinions over time.

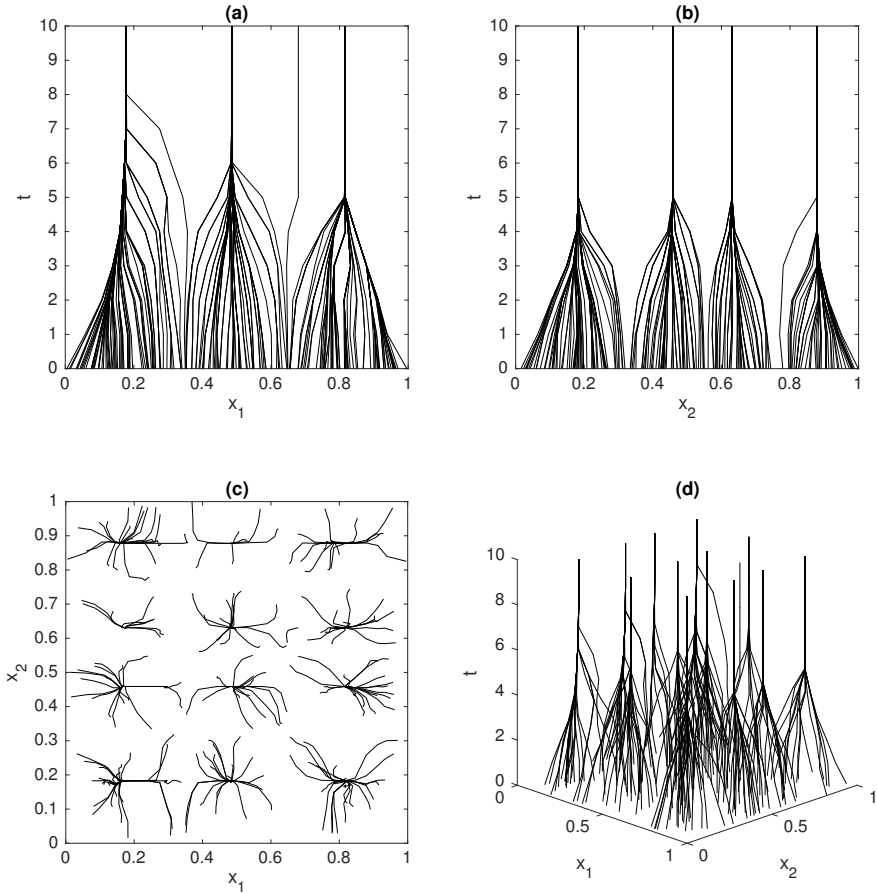

**Figure 2:** Evolution of opinions when issues are independent, resulting in 12 final clusters. (a) and (b) show the evolution of $x_1$ and $x_2$ respectively, (c) shows the trace in both dimensions, and (d) shows the evolution of both over time. Simulation parameters used were $n = 200$, $m = 2$, and $\epsilon_k = 0.1$ for all $k$.

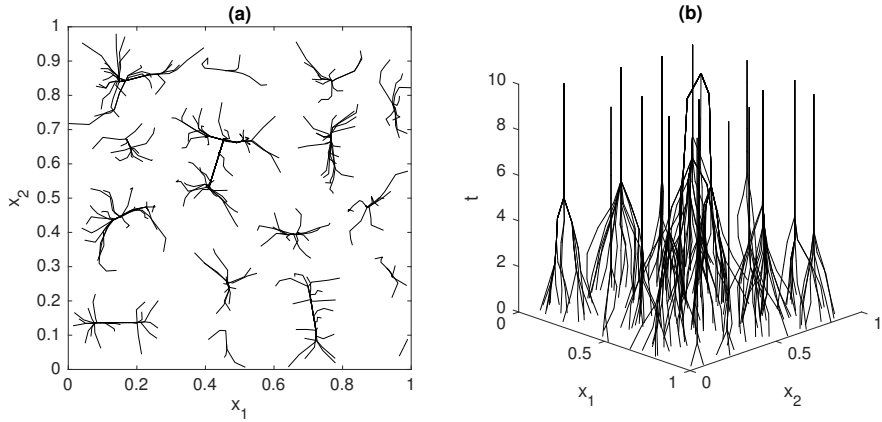

**Figure 3:** Evolution of opinions under exclusivist interaction, resulting in 16 final clusters. (a) shows the trace in both dimensions, and (b) shows the evolution over time. Simulation parameters used were $n = 200$, $m = 2$, and $\epsilon_k = 0.1$ for all $k$.

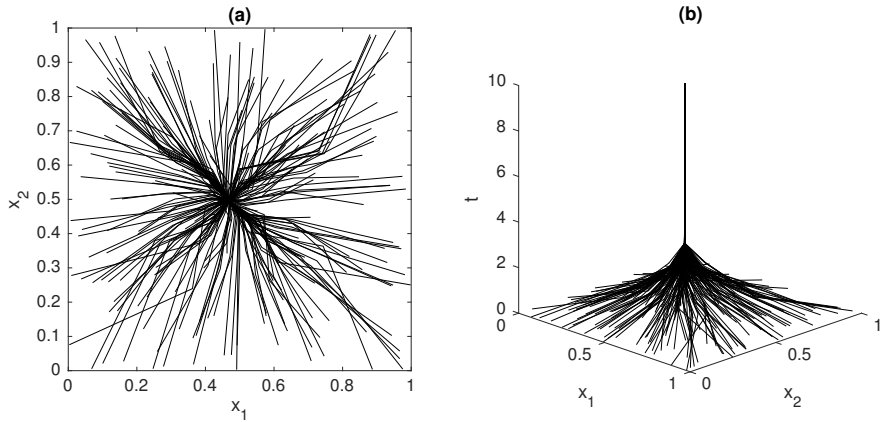

**Figure 4:** Evolution of opinions under inclusivist interaction, resulting in 1 final cluster. (a) shows the trace in both dimensions, and (b) shows the evolution over time. Simulation parameters used were $n = 200$, $m = 2$, and $\epsilon_k = 0.1$ for all $k$.

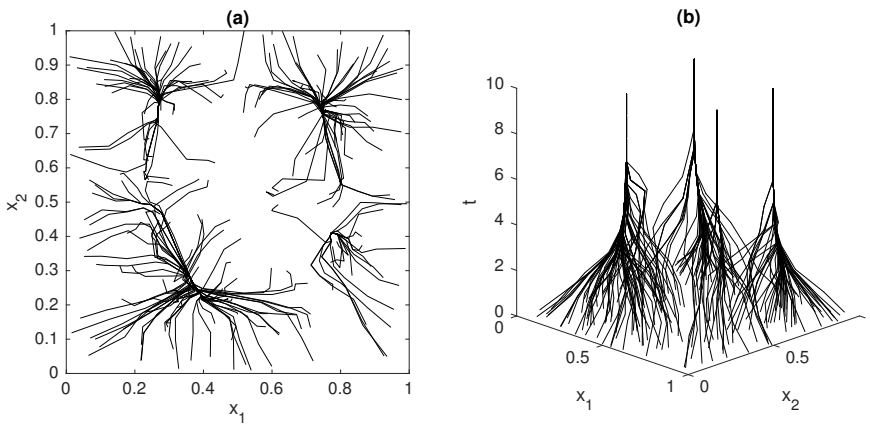

**Figure 5:** Evolution of opinions under the general model of inclusivity, resulting in 4 final clusters. (a) shows the trace in both dimensions, and (b) shows the evolution over time. Simulation parameters used were $n = 200$, $m = 2$, $\epsilon_k = 0.1$ for all $k$, and $\alpha_k = 2.5$ for all $k$.

The case where issues are independent from each other is depicted in Figure 2. As can be seen from Figures 2(a) and (b), opinions on each issue evolve independently and converge to several non-interacting clusters, such that in two dimensions, the clusters form a rectangular lattice. The total number of opinion clusters ($= 12$) is hence just the product of the number of clusters for each issue (3 and 4 for the first and second issue respectively). As a result, even though opinion fragmentation for each issue is limited, when both issues are considered together, fragmentation is much more severe. This can be seen in Figures 2(c) and (d).

Figures 3 and 4 show the evolution of opinions under exclusivist and inclusivist interaction respectively. As expected from the smaller neighborhoods of acceptance, fragmentation is severe in the exclusivist case, with 16 final clusters emerging, and no lattice arrangement present as in the independent case. This lack of regular structure is because issues are no longer considered independently, and so if fragmentation emerges, it emerges across all issues simultaneously (vice versa for consensus). For the inclusivist case, fragmentation is non-existent – total consensus emerges very rapidly instead, in line with the large neighborhoods of acceptance.

Figure 5 shows the evolution of opinions under generalized inclusivity, with degree of inclusivity $\alpha_k$ set to 2.5 for all $k$. Four non-interacting opinion clusters emerge. As expected, this lies between the two extremes generated by the exclusivist and inclusivist scenarios. Again it can be seen that the clusters are not arranged in a lattice. Rather, because fragmentation occurs across both issues at once, there is some correlation between an individual's final opinion on the first issue and the seccond issue. For example, if individuals are far to the left on the first issue, they are more likely to be in the larger bottom-left cluster in Figure 5(a), and thus are more likely to be near the bottom on the second issue.

To better quantify the effect of inclusivity and dimensionality upon consensus and polarization, the average number of final opinion clusters $n_c$ was computed via simulation for a range of dimensions and degrees of inclusivity. The number of clusters $n_c$ is used as a measure of the degree of opinion fragmentation (correspondingly, an inverse measure of the degree of consensus). The results of these simulations are shown in Table 1 and depicted in Figure 6. All other parameters were kept constant at the following values: $n = 1000$ agents, $T = 50$ timesteps, a constant latitude of acceptance $\epsilon_k = 0.1$ for all dimensions. For each simulation, the degree of inclusivity was equal across all dimensions ($\alpha_k = \alpha$ for all dimensions $k$), and for each degree of inclusivity $\alpha$ and dimension $m$, the data was averaged over 100 trials. It is important to note that $\alpha = 1.0$ (bolded in Table 1) corresponds to exclusivism, and that $\alpha = 10$ (also bold) corresponds to inclusivism (because when $\epsilon_k = 0.1$, $\alpha = 10$ has the same effect as $\alpha = \infty$).

|  | Dimensions, $m$ | | | | | | | | | |
|---|---|---|---|---|---|---|---|---|---|---|
| Inclusivity $\alpha$ | 1 | 2 | 3 | 4 | 5 | 6 | 7 | 8 | 9 | 10 |
| **1.0** | 3.9 | 14.21 | 61.61 | 441.93 | 872.8 | 976.7 | 995.95 | 999.09 | 999.89 | 999.97 |
| 2.5 | 3.9 | 2.71 | 1.92 | 1.58 | 3.41 | 17.38 | 94.92 | 318.48 | 634.71 | 850.02 |
| 3.25 | 3.9 | 1.01 | 1 | 1 | 1 | 1.13 | 2.57 | 11.41 | 39.21 | 113.27 |
| 4.0 | 3.9 | 1 | 1 | 1 | 1 | 1 | 1.02 | 1.11 | 1.74 | 3.72 |
| 4.75 | 3.9 | 1 | 1 | 1 | 1 | 1 | 1 | 1 | 1.03 | 1.01 |
| 5.5 | 3.9 | 1 | 1 | 1 | 1 | 1 | 1 | 1 | 1 | 1 |
| 7.0 | 3.9 | 1 | 1 | 1 | 1 | 1 | 1 | 1 | 1 | 1 |
| 8.5 | 3.9 | 1 | 1 | 1 | 1 | 1 | 1 | 1 | 1 | 1 |
| **10** | 3.9 | 1 | 1 | 1 | 1 | 1 | 1 | 1 | 1 | 1 |

**Table 1:** Effect of inclusivity and dimensionality on the number of clusters $n_c$ after $T = 50$ timesteps. Other parameters were $n = 1000$, $\epsilon_k = 0.1$ for all $k$. Results were averaged over 100 trials.

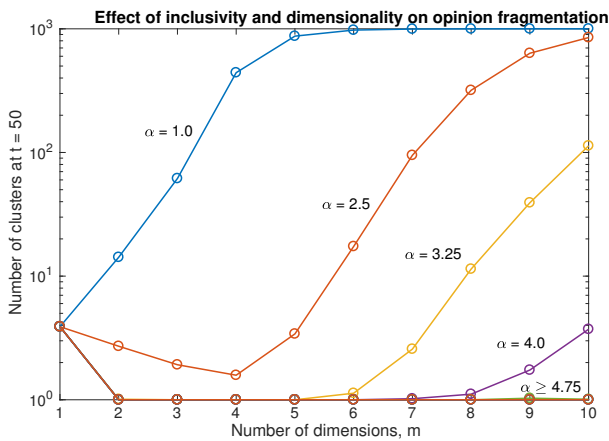

**Figure 6:** Effect of inclusivity and dimensionality on the number of clusters $n_c$ after $T = 50$ timesteps. For $\alpha \geq 4.5$, $n_c$ was equal or very close to 1 for all $m \geq 2$, and hence the trend lines appear to overlap.

In general, it can be seen that the degree of fragmentation increases as the number of issues $m$ increases, but decreases as the degree of inclusivity $\alpha$ increases. Both of these trends are as predicted. Furthermore, as $m$ increases, the value to which $\alpha$ has to reach to maintain absolute consensus ($n_c = 1$) also increases. In other words, greater inclusivity is required to ensure solidarity and cohesion in cross-issue interactions as more issues come under consideration.

A minor deviation from these general trends is that for $\alpha > 1.0$, the number of clusters drops from $m = 1$ to $m = 2$. This is simply because $\alpha$ has no effect in the one-dimensional case, but expands the neighborhood of acceptance for all $m \geq 2$. Thus, the amount of fragmentation is constant for $m = 1$ regardless of $\alpha$, but decreases with alpha once $m \geq 2$.

Interestingly, with the given parameters, there is a point at which greater inclusivity brings no further consensus — when $\alpha = 4.75$, consensus is effectively ensured for up to $m = 10$ dimensions, and larger values of $\alpha$ show minimal improvement (though they might show more improvement when $m$ grows even larger). This is a promising result for the promotion of cross-issue solidarity, because it suggests that a limited amount of inclusivity is enough for cohesion in practical contexts where the number of issues is not arbitrarily high.

A similar kind of saturation also occurs with the degree of opinion fragmentation. This effect is most clearly seen in the exclusivist case ($\alpha = 1.0$). As $m$ increases, $n_c$ eventually reaches the maximum of $n = 1000$ – i.e., just about every individual is isolated in a non-interacting cluster. This extreme amount of fragmentation is because agreement on too many issues has to occur before these exclusivist individuals interact with one another. Clearly, exclusivism should be avoided by those who wish to promote cross-issue cohesion.

To further validate the trends described, analytical derivations were also performed to determine the likelihood of consensus as a function of $\alpha$. In accordance with our simulations, the two-agent likelihood of consensus increases with $m$ when $\alpha$ is high, and decreases when $\alpha$ is low. Indeed, when $\epsilon = 0.1$, there is a threshold of inclusivity $\alpha^* = 5$ above which the likelihood of consensus always increases with $m$. This somewhat counter-intuitive result readily explains the observation that consensus seems to saturate when $\alpha \geq 4.75$. Full details of these derivations can be found in the Supplementary Information.

## 4.2 Effects of inclusivity on convergence to the truth

Another set of simulations was performed to investigate the dynamics of opinion evolution in the presence of truth, modelled as a virtual agent with opinion $T$. Given that the extremity of the truth with respect to the average person's opinion might influence the degree of convergence toward the truth, three scenarios were considered — central truth (close to the average opinion), moderate truth (somewhat removed from the average opinion), and extreme truth (distant from the average opinion). Since the opinions of the simulated population were initially distributed uniformly at random across the opinion space $[0, 1]^m$ (hence having an expected average opinion of $(0.5_{\times m})$, i.e. 0.5 repeated $m$ times), we chose the values of $T = (0.45_{\times m})$, $(0.30_{\times m})$, $(0.15_{\times m})$ to correspond to the respective scenarios.

These scenarios were simulated for both $m = 2$ and $m = 3$ dimensions, and with varying degrees of inclusivity $\alpha$. All simulations were run for $T = 100$ timesteps, and used a smaller number of $n = 50$ individuals to avoid "drowning out" the truth. (Recall that the truth has the influence of a single individual in our model, though a simple modification of the model can allow this influence to be weighted stronger or weaker instead.) For each simulation, the number of final opinion clusters $n_c$ was recorded, as was the fraction of the population that eventually converged to the truth, $f_T$. The results for each set of parameters were averaged over 100 trials.

Figure 7 shows two sample runs for each of the three scenarios when $m = 2$, with the first sample run simulated with $\alpha = 2.5$ (low inclusivity) and the second sample run simulated with $\alpha = 7.0$ (high inclusivity). In line with the results presented earlier, a higher degree of inclusivity results in greater consensus. Greater consensus, however, does not always coincide with a greater degree of convergence to the truth (as measured by $f_T$). Instead, convergence to the truth depends on both how extreme the truth is, and the degree of inclusivity $\alpha$.

Under high inclusivity ($\alpha = 7.0$), most of the population converges to the truth if it is central. This is because in the absence of truth, the expected point of consensus is at the initial average opinion $\bar{x}$ (in this case, the center of the opinion space), which is near the truth. As can be seen in Figure 7b, the population rapidly converges towards $\bar{x}$. Following that, the population slowly approaches the truth. These dynamics occur because the influence of the truth is relatively small, and can hence only exert so much pull on the rest of the population. Nonetheless, the truth $T$ is within the neighborhood of acceptance of $\bar{x}$, and so the population converges to it after a while. On the other hand, if the truth is more extreme, it will fall outside the neighborhood of acceptance of $\bar{x}$, such that under high inclusivity, the population does not converge to the truth at all. This can be seen in Figures 7d and 7f. Rapid convergence towards $\bar{x}$ still occurs, which means that even individuals initially close to $T$ end up moving away so quickly that they are hardly influenced by the truth at all. Once they move away, $T$ is no longer within their neighborhood of acceptance, and so they never converge to the truth.

When inclusivity is low ($\alpha = 2.5$), the dynamics are markedly different. Given a central truth (Figure 7a), the fraction that converges toward it is much smaller than the high inclusivity case. This is because individuals with initially extreme opinions end up converging to extreme clusters instead, and hence are too far from the truth to "listen" to it. While this might seem like a negative outcome, it is precisely this potential to form extreme clusters that allows for convergence to more extreme truths. As shown in Figures 7c and 7e, significant fractions of the population still converge to truth when it is moderate or extreme, a sharp contrast to the high inclusivity case, where the entire population ignores more extreme truths. It can also be seen that smaller clusters converge more rapidly to the truth (compare for e.g. Figures 7e and 7b). This is because the truth has greater relative influence over smaller groups of individuals, whereas in large clusters, it has to compete with the voices of a lot more individuals — a counterintuitive benefit of opinion fragmentation.

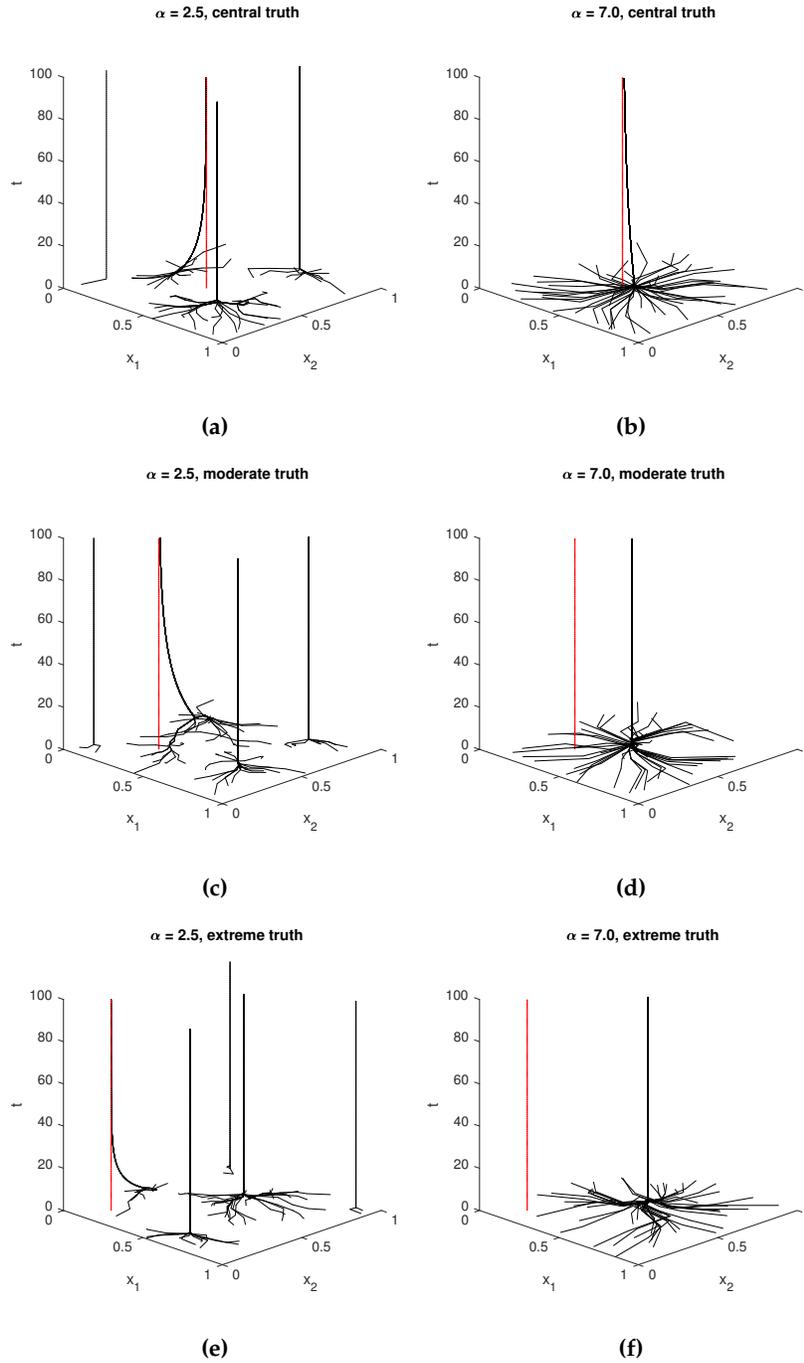

**Figure 7:** Evolution of opinions under varying inclusivity and extremity of truth, where the $z$ axis is time. The trace of the truth $T$ is depicted by a red dashed line. For all simulations shown, $n = 50, T = 100, m = 2$, and $\epsilon_k = 0.1$ for all $k$.

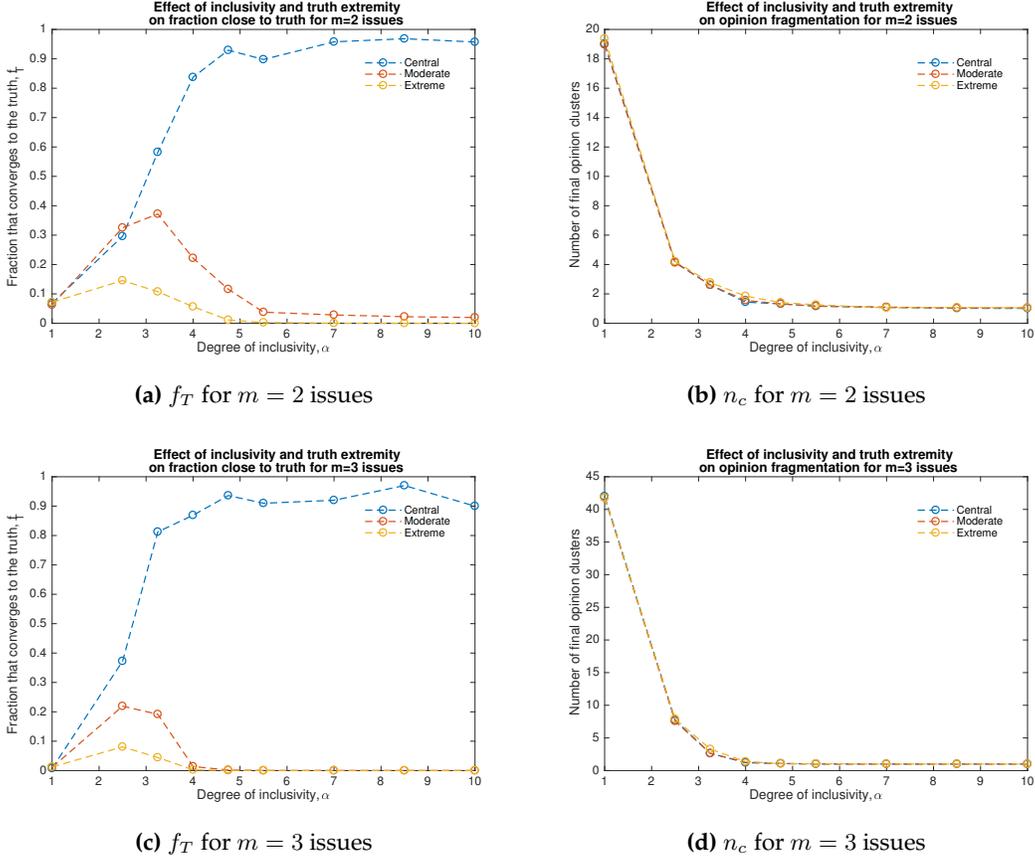

**Figure 8:** Effects of inclusivity and truth extremity on the fraction that converges to the truth $f_T$ and the number of clusters $n_c$ after $T = 100$ timesteps. Other parameters were $n = 50$, $\epsilon_k = 0.1$ for all $k$. Results were averaged over 100 trials.

Figure 8 shows the aggregate results for the effects of inclusivity and truth extremity for $m = 2$ and 3 issues, demonstrating that trends described above hold more generally. As Figures 8a and 8c show, $f_T$, the fraction that converges to the truth, increases with $\alpha$ when the truth is central. When the truth is moderate or extreme however, $f_T$ first increases then peaks at a lower value of $\alpha$, before dropping close to zero and remaining there for $\alpha \geq 4.75$. Both overly high and overly low $\alpha$ result in lack of convergence to the truth, but for different reasons. The case of high inclusivity has been described. For overly low inclusivity, lack of truth convergence is due to the large degree of fragmentation, such that even if some fragments end up close enough to $T$ to converge towards it, their size is still very small. Between these two extremes, there is a 'sweet spot' where $f_T$ is maximal. At this point, inclusivity is low enough to prevent groupthink (i.e. fallacious consensus at the average opinion), but high enough that large fragments of the population eventually converge to the truth.

In most cases, a more extreme truth results in smaller $f_T$, regardless of $\alpha$. This is to be expected, because less individuals are close to truth when it is further from the center of the initial opinion distribution, and so less individuals end up converging towards it. On the other hand, Figures 8b and 8d show that truth extremity has minimal impact on the degree of fragmentation, as measured by $n_c$. In agreement with the previous section's results, $n_c$ drops rapidly to 1 as $\alpha$ increases. The effects of dimensionality are also similar — there is higher fragmentation (higher $n_c$) for $m = 3$, which in turn explains the lower $f_T$ for $m = 3$ when the truth is not central.

## 5  Discussion and conclusion

As hypothesized in our presentation of the models, the results show that greater inclusivity leads to a higher degree of cohesion and consensus, whereas a greater number of issues increases the degree of fragmentation. While this is a straightforward enough result, it formalizes an important insight for actors in the public sphere who wish to promote cross-issue solidarity. As more issues come under collective debate, the risk of disagreement increases, and in order to mitigate this risk, participants in the conversation have to also increase their degree of inclusivity.

On this note, it is crucial to distinguish between 'inclusivity' as we have quantified it here, and the slightly different concept of 'open-mindedness'. Open-mindedness can be understood as the general willingness to listen to dissimilar viewpoints, and in bounded-confidence models, it is captured by $\epsilon$, the latitude of acceptance. Inclusivity, on the other hand, is one's willingness to hear another out *when already in agreement on something else*, and is captured in our models by $\alpha$. Becoming more inclusive is thus, psychologically speaking, a lower bar. It does not require greater acceptance of arbitrary opinions, only the opinions of those with whom one already shares something in common. This matters, because while it may be difficult to increase the open-mindedness of the average person, encouraging inclusivity towards those already in partial agreement is likely to be considerably easier.

How might inclusivity be encouraged? One possibility is to simply raise awareness about how unrealistic it is to expect simultaneous agreement on more than a few issues, and how insisting upon such agreement (i.e. engaging in exclusivist interaction) can lead to severe fragmentation. Returning to the motivating example of intersectional activism, it is just this potential for fragmentation that critics often lament.[8,10] By formalizing and demonstrating the effects of exclusivism, the results presented here may aid in dissuading such practices. Instead, our results offer the inclusivist mode of interaction as a more productive alternative for multi-issue dialogue. As defenders of intersectional movements argue, inclusivity is what such movements look like at their best — they 'draw people in' rather than 'call people out', using shared commitments to promote further dialogue without erasing differences.[11,13] Although one's degree of inclusivity might ultimately be limited, as our results show, limited inclusivity is sufficient for consensus. In fact, with enough inclusivity, consensus can become even *more* likely as more issues get involved — as participants discuss more issues, they discover more potential areas of agreement, which can then promote engagement on other issues that they differ on.

As suggested in the introduction however, consensus may not always lead societies or movements to the truth. Indeed, as our results demonstrate, consensus and convergence to the truth can sometimes be quite at odds. In particular, the more extreme the truth is, the more likely that any consensus achieved will be far removed from it. Furthermore, since greater inclusivity promotes consensus, too much inclusivity, while minimizing fragmentation, may lead to extreme truths being wholly ignored.

Intriguingly and also importantly, whether this dilemma is concerning may depend on where one's opinion already lies. This is because people tend to assume that their own opinion is relatively close to the truth (otherwise, they would not have that opinion). For centrists, who think the truth is roughly in between most people's opinions (and therefore have central opinions themselves), greater inclusivity is a non-issue, because it promotes consensus at the center, and thus eventual convergence to where they think the truth is. For moderates, limited inclusivity is helpful, but too much will result in convergence of opinions towards a central position that (at least initially) they think misguided. For radicals, who think the truth is extreme, the dilemma between truth and consensus is the most pressing of all. Too much inclusivity, and they might end up giving up on (what they believe to be) hard-won truths. Too little inclusivity, and the fragmentation might be too strong ever induce social change.

What this implies for the simultaneous pursuit of consensus and truth is uncertain, and further research is needed. Tentatively, we suggest taking into account the different strategies of inclusivity that actors might employ depending on whether they are centrists, moderates, or radicals, along with the fact that inclusivity can change with time. One possible route to consensus at an extreme truth, for example, might be for nearby radical actors to limit their inclusivity at first, gathering a strong enough pool of like-minded individuals before increasing their inclusivity and using their greater numbers to draw moderates and centrists into their cluster. All this assumes, of course, that radicals on other fringes of the opinion space do not try the same approach.

Nonetheless, it remains the case across scenarios that too little inclusivity can only result in unproductive fragmentation — so much fragmentation that even if pockets of individuals converge to the truth, their number will be minuscule. However one balances the trade-off between truth and consensus then, it is still advisable to avoid exclusivist interaction. If not, people will neither listen to the truth, nor will they listen to each other.

Collectively, these findings promise great theoretical and practical significance, and are of direct relevance to contemporary political debates. By formalizing intuitive descriptions of how people might interact with the truth and engage when discussing multiple issues, our work allows opinion dynamics to capture a wide new range of phenomena. The conclusions from our models can be used to guide how political campaigns (and other relevant processes) run. It also helps refocus the debate on divisive politics by going beyond simplistic assumptions about multi-issue dialogue, and turning attention towards the inclusivity of cross-issue interactions instead — a move we believe will be of substantial benefit to observers and organizers alike.

# References

[1] U. Friedman, "What If the 'Populist Wave' Is Just Political Fragmentation?" *The Atlantic*, Mar. 2017. [Online]. Available: https://www.theatlantic.com/international/archive/2017/03/dutch-election-wilders-populism/519813/

[2] Staff, "The party declines," *The Economist*, Mar. 2016. [Online]. Available: http://www.economist.com/node/21694006

[3] S. Kutchinsky, "Yanis Varoufakis: After Donald Trump's awful victory, the left must be more ambitious," *New Statesman*, Nov. 2016. [Online]. Available: http://www.newstatesman.com/politics/brexit/2016/11/yanis-varoufakis-after-donald-trump-s-awful-victory-left-must-be-more

[4] D. Miliband, "Full Transcript | David Miliband | Speech on the European left | LSE | 8 March 2011," *New Statesman*, Mar. 2011. [Online]. Available: http://www.newstatesman.com/uk-politics/2011/03/centre-parties-social

[5] C. Cochrane, "The asymmetrical structure of left/right disagreement: Left-wing coherence and right-wing fragmentation in comparative party policy," *Party Politics*, vol. 19, no. 1, pp. 104–121, 2013.

[6] J. Steinhauer, "Without Obama as a Unifier, Republicans Are Fragmented," *The New York Times*, Jun. 2017. [Online]. Available: https://www.nytimes.com/2017/06/06/us/politics/without-obama-as-a-unifier-republicans-are-fragmented.html

[7] F. Stockman, "Women's March on Washington Opens Contentious Dialogues About Race," *The New York Times*, Jan. 2017. [Online]. Available: https://www.nytimes.com/2017/01/09/us/womens-march-on-washington-opens-contentious-dialogues-about-race.html

[8] D. Linker, "Liberals are drunk on a political poison called intersectionality," *The Week*, Jan. 2017. [Online]. Available: http://theweek.com/articles/672265/liberals-are-drunk-political-poison-called-intersectionality

[9] H. Wilhelm, "Women's March Morphs into Intersectional Torture Chamber," *National Review*, Jan. 2017. [Online]. Available: http://www.nationalreview.com/article/443741/womens-march-feminists-oppose-donald-trump-struggle-agree-how

[10] H. Lewis, "The uses and abuses of intersectionality," *New Statesman*, Feb. 2014. [Online]. Available: http://www.newstatesman.com/helen-lewis/2014/02/uses-and-abuses-intersectionality

[11] J. Desmond-Harris, "To understand the Women's March on Washington, you need to understand intersectional feminism," *Vox*, Jan. 2017. [Online]. Available: http://www.vox.com/identities/2017/1/17/14267766/womens-march-on-washington-inauguration-trump-feminism-intersectionaltiy-race-class

[12] M. Goldberg, "Democratic Politics Have to Be Identity Politics," *Slate*, Nov. 2016. [Online]. Available: http://www.slate.com/articles/news_and_politics/politics/2016/11/democratic_politics_have_to_be_identity_politics.html

[13] D. Roberts and S. Jesudason, "Movement intersectionality," *Du Bois Review: Social Science Research on Race*, vol. 10, no. 02, pp. 313–328, 2013.

[14] J.-H. Zhao, Z. Hai-Jun, and Y.-Y. Liu, "Inducing effect on the percolation transition in complex networks," *Nature Communications*, vol. 4, p. ncomms3412, 2013.

[15] M. Domenico, C. Granell, M. Porter, and A. Arenas, "The physics of spreading processes in multilayer networks," *Nature Physics*, vol. 12, p. nphys3865, 2016.

[16] G. Yan, G. Tsekenis, B. Barzel, J.-J. Slotine, Y.-Y. Liu, and A.-L. Barabási, "Spectrum of controlling and observing complex networks," *Nature Physics*, vol. 11, p. nphys3422, 2015.

[17] F. Xiong, Y. Liu, and J. Cheng, "Modeling and predicting opinion formation with trust propagation in online social networks," *Communications in Nonlinear Science and Numerical Simulation*, vol. 44, pp. 513–524, 2017.

[18] Y. Shang, "An agent based model for opinion dynamics with random confidence threshold," *Communications in Nonlinear Science and Numerical Simulation*, vol. 19, no. 10, pp. 3766–3777, 2014.

[19] H. Chau, C. Wong, F. Chow, and C.-H. F. Fung, "Social judgment theory based model on opinion formation, polarization and evolution," *Physica A: Statistical Mechanics and its Applications*, vol. 415, pp. 133–140, 2014.

[20] A. Nowak, J. Szamrej, and B. Latané, "From private attitude to public opinion: A dynamic theory of social impact." *Psychological Review*, vol. 97, no. 3, p. 362, 1990.

[21] L. Salzarulo, "A continuous opinion dynamics model based on the principle of meta-contrast," *Journal of Artificial Societies and Social Simulation*, vol. 9, no. 1, 2006.

[22] M. Moussaïd, J. E. Kämmer, P. P. Analytis, and H. Neth, "Social influence and the collective dynamics of opinion formation," *PloS one*, vol. 8, no. 11, p. e78433, 2013.

[23] R. Hegselmann, U. Krause *et al.*, "Opinion dynamics and bounded confidence models, analysis, and simulation," *Journal of Artificial Societies and Social Simulation*, vol. 5, no. 3, 2002.

[24] G. Weisbuch, G. Deffuant, F. Amblard, and J.-P. Nadal, "Meet, discuss, and segregate!" *Complexity*, vol. 7, no. 3, pp. 55–63, 2002.

[25] G. Deffuant, D. Neau, F. Amblard, and G. Weisbuch, "Mixing beliefs among interacting agents," *Advances in Complex Systems*, vol. 3, no. 01n04, pp. 87–98, 2000.

[26] K. Sznajd-Weron and J. Sznajd, "Opinion evolution in closed community," *International Journal of Modern Physics C*, vol. 11, no. 06, pp. 1157–1165, 2000.

[27] W. Ren, R. W. Beard, and E. M. Atkins, "A survey of consensus problems in multi-agent coordination," in *Proceedings of the 2005, American Control Conference, 2005.* IEEE, 2005, pp. 1859–1864.

[28] C. Wang, Z. X. Tan, Y. Ye, L. Wang, K. H. Cheong, and N.-g. Xie, "A rumor spreading model based on information entropy," *Scientific Reports*, vol. 7, p. 9615, 2017.

[29] P. Grindrod and D. J. Higham, "A dynamical systems view of network centrality," *Proc. R. Soc. A*, vol. 470, p. 20130835, 2014.

[30] B. Voorhees and A. Murray, "Fixation probabilities for simple digraphs," *Proc. R. Soc. A*, vol. 469, p. 20120676, 2012.

[31] A. C. Martins, C. d. B. Pereira, and R. Vicente, "An opinion dynamics model for the diffusion of innovations," *Physica A: Statistical Mechanics and its Applications*, vol. 388, no. 15, pp. 3225–3232, 2009.

[32] M. Sherif and C. I. Hovland, *Social judgment: Assimilation and contrast effects in communication and attitude change.* Yale Univer. Press, 1961.

[33] J. Lorenz, "Continuous opinion dynamics under bounded confidence: A survey," *International Journal of Modern Physics C*, vol. 18, no. 12, pp. 1819–1838, 2007.

[34] ——, "Continuous opinion dynamics of multidimensional allocation problems under bounded confidence: More dimensions lead to better chances for consensus," *arXiv preprint arXiv:0708.2923*, 2007.

[35] ——, "Fostering consensus in multidimensional continuous opinion dynamics under bounded confidence," in *Managing complexity: insights, concepts, applications.* Springer, 2008, pp. 321–334.

[36] S. Huet, G. Deffuant, and W. Jager, "A rejection mechanism in 2d bounded confidence provides more conformity," *Advances in Complex Systems*, vol. 11, no. 04, pp. 529–549, 2008.

[37] D. Baldassarri and P. Bearman, "Dynamics of political polarization," *American sociological review*, vol. 72, no. 5, pp. 784–811, 2007.

[38] D. N. Walton, "Why is the'ad populum'a fallacy?" *Philosophy & Rhetoric*, pp. 264–278, 1980.

[39] R. Bellamy and M. Hollis, "Consensus, neutrality and compromise," *Critical Review of International Social and Political Philosophy*, vol. 1, no. 3, pp. 54–78, 1998.